\documentstyle[11pt,russian]{article}
\begin{document}
\begin{center}{\bf GEOMETRISATION OF ELECTROMAGNETIC FIELD AND TOPOLOGICAL
INTERPRETATION OF QUANTUM FORMALISM}\end{center}
\begin{center}O.A.OLKHOV\\
{\it Semenov Institute of Chemical Physics, Russian Academy of 
Sciences, Moscow}\\
{\it e-mail: olkhov@center.chph.ras.ru}\end{center}
\par\bigskip

\noindent     A new concept of geometrization of electromagnetic field is 
proposed. Instead of the concept of extended field and its point 
sources, the interacting Maxwellian and Dirac electron--positron 
fields are considered as a microscopic unified closed connected nonmetrized 
space--time 4-manifold. Within this approach, the Dirac equation 
proves to be a group-theoretic relation that accounts for the 
topological and metric properties of this manifold. The Dirac 
spinors serve as basis functions of its fundamental group 
representation, while the tensor components of electromagnetic 
field prove to be the components of a curvature tensor of the relevant covering 
space. A basic 
distinction of the suggested approach from the geometrization of 
gravitational field in general relativity is that, first, not 
only the field is geometrized but also are its microscopic 
sources and, second, the field and its sources are treated not as 
a metrized Riemannian space--time but as a nonmetrized space--
time manifold. A possibility to geometrize weak interaction is 
also discussed.
\par\bigskip

\noindent{\bf Introduction}
\par\medskip
     The possibility of representing physical interaction as a 
distortion of pseudo-Euclidean space--time metric was first 
demonstrated in general relativity. However, further attempts at 
including other interactions and, primarily, electromagnetic 
interaction into the analogous geometrical approach have failed. 
It is shown in this work that electromagnetic field and its 
microscopic sources can be represented as a unified nonmetrized 
space--time topological manifold. Some of the preliminary results 
of this work have already been published and delivered at 
conferences [1].
     Let us first use an analogy with the theory of general 
relativity to outline the main results of this work. It is 
demonstrated in general relativity that the classical equations 
of motion in gravitational field can be interpreted as equations 
for the shortest distances between two points of a curved 
Riemannian space--time. It is shown in this work that the Maxwell 
and Dirac equations for the interacting electromagnetic field and 
its quantum microscopic sources (electron--positron fields) can 
be interpreted as a group--theoretic relations describing a closed 
connected nonorientable nonmetrized space--time 4-manifold whose 
topological invariants are the observable physical 
characteristics of the system considered. Within above approach the 
possible equation for 
neutral Fermi particles interacting via weak 
field generated by them is discussed. This interaction violates the space and 
time symmetry. The possibility of interpreting this equation as 
equation describing, in the classical limit, weak interaction of 
neutrinos is discussed.

     The interpretation of the microscopic field sources as 
extended geometrical objects forming, together with field, a 
unified topological 4-manifold requires a 
new geometrical interpretation of the 
mathematical apparatus of quantum mechanics. This new 
interpretation, above all, should not contradict the traditional 
interpretation at least in those cases where the latter approach 
has confirmed its effectiveness experimentally. For this reason, 
before passing to the main part of this work, it is worthwhile to 
show that the concept of the quantum object as a special 
nonmetrized topological manifold is not contradictory to the 
fundamental positions of the conventional interpretation of 
quantum mechanics.

     1. The de Broglie's original idea of "travelling wave," 
which is specified for a free particle in the wave--corpuscular 
duality concept by the function
$$\exp i(\omega t-{\bf kr})=\exp i\left(\frac {E}{\hbar}t-\frac {\bf p}{\hbar}
{\bf r}\right), \eqno (1)$$
underlies the traditional interpretation of quantum mechanics. 
However, Eq. (1) can be recast in the form that allows a 
different, not wave, interpretation. Let us write Eq. (1) in the 
"relativistically symmetric" form
$$\exp i\left(\frac{x_1}{l_1}-\frac{x_2}{l_2}-\frac{x_3}{l_3}-
\frac{x_4}{l_4}\right),\eqno (2)$$
where $x_1=ct, x_2=x, x_3=y, x_4=z$ and $l_1=2\pi \hbar c/E, l_2=2\pi \hbar/p_x,
l_3=2\pi \hbar/p_y, l_4=2\pi \hbar/p_z$.

     In the group theory, function (2) plays the role of a 
representation of the cyclic Abelian group with four generators 
$l_1^{-1},l_2^{-1},l_3^{-1}$ and $l_4^{-1}$ [2]. 
Because of this, the object described by this 
function can be interpreted not only as possessing wave 
properties but also as a geometrical object possessing a certain 
symmetry.

     2. The notion of a point particle has no physical meaning in 
the relativistic quantum mechanics. Since a change in momentum 
during measurement cannot be as large as one likes because of the 
limiting speed of light, the uncertainty in measuring coordinates 
cannot be as small as is wished [3]. This does not contradict the 
treatment of electromagnetic field and its microscopic sources as 
an extended space--time manifold.

     3. The traditional quantum mechanical formalism is based on 
the statistical probabilistic description. This does not 
contradict the description of a quantum object as a nonmetrized 
topological manifold that is "chaotic" by its very topological 
nature.

     4. The suggested geometrical interpretation of the quantum 
formalism does not involve any "hidden" parameters, whose 
incorporation in any new possible interpretation of quantum 
mechanics is forbidden by the relevant theorems [4, 5].

     In closing this section, two fundamental distinctions 
between the suggested topological approach and the geometrization 
of gravitational field in general relativity should be 
emphasized.
     1. The field sources (masses) in general relativity are 
considered as nongeometrized point objects. Only the extended 
field is the object of geometrization. In this work, the 
microscopic field sources are treated as extended objects which 
form, together with field, a single unified geometrical object.
     2. The gravitational field in general relativity is put in 
correspondence with a 
metrized topological manifold, namely, with a curved Riemannian space 
that has a certain form in every 
instant of time. In this work, the field and its sources are put 
in correspondence with an essentially different geometrical 
object, namely, with a nonmetrized manifold whose metric is 
undetermined and, hence, which has no definite form and is 
characterized only by its topological invariants.
\par\medskip
\noindent     {\bf Topological derivation of Dirac equation}
\par\medskip

      It is my purpose to prove that the topological 
characteristics of a certain closed connected 4-manifold are 
encoded in a certain way in the equations for a classical field 
interacting with its microscopic sources (electrons and 
positrons). These equations (Maxwell and Dirac equations) have 
the form [3]
$$i\gamma_1\left (\frac{\partial}{\partial x_1}-ieA_1 \right)\psi-
\sum_{\alpha =2}^4 i\gamma_\alpha \left(\frac
{\partial}{\partial x_\alpha}-ieA_\alpha \right)\psi=m\psi,\eqno (3)$$
$$F_{kl}=\frac{\partial A_k}{\partial x_l}-\frac{\partial A_l}{\partial x_k},
\eqno (4)$$
$$\sum_{i=1}^4 \frac{\partial F_{ik}}{\partial x_i}=j_k,\quad
j_k=e\psi*\gamma_1\gamma_k\psi.\eqno (5)$$ 
Here $\hbar=c=1,\quad x_1=t,\quad x_2=x,\quad x_3=y,\quad x_4=z,\quad F_{kl}$ 
is the electromagnetic field tensor, $A_k$ is 
the 4-potential, $\gamma_k$ are the well-known Dirac matrices, $\psi$ is the 
Dirac bispinor, and $m$ and $e$ are electron mass and charge, 
respectively.

To my knowledge, the topological manifolds are not 
identified by the differential equations. This, in particular, 
differentiates the suggested approach from general relativity, 
for which the mathematical apparatus (e.g., equations for 
geodesics) had well been elaborated by the time the theory 
emerged. For this reason, I will first demonstrate, by a simple 
example, how can the differential equations describe, in 
principle, the topological characteristics of a manifold.

Let us consider the simplest closed connected nonmetrized 
manifold, namely, a one-\\ dimensional $S^1$ manifold homeomorphic to 
(equivalent to) ring. It is equivalent in the sense that it can 
be represented as any of the objects obtained from ring by its 
deformation without discontinuities. A fundamental group of 
different classes of closed paths starting and ending at the same 
point of the manifold is one of the topological invariants of any 
connected manifold [6]. Free cyclic group isomorphous with the $Z$ 
group (integer group, a topological invariant of the manifold 
considered) is the fundamental group of $S^1$ [7].
     In turn, the Z group is isomorphous with the group of 
parallel translations along a straight line with one generator 
(the line is called the covering space of our manifold). Denote 
the generator length by $l$. Note that the operator $T_{lx}$
$$T_{lx}=\left(\frac {il}{2\pi} \right)\frac {d}{dx}\eqno (6)$$
can be regarded as a representation of the above group with a 
basis defined by the function $\varphi_l(x)$
$$\varphi_l(x)=\exp \left(-2\pi i \frac {x}{l} \right).\eqno (7)$$
Indeed,
$$\varphi'_l=\varphi_l(x+l)=T_{lx}\varphi_l(x).\eqno (8)$$

     Thus, both the fundamental group (a topological invariant of 
the manifold considered) and the constraint (metric 
characteristic of this manifold) on the length of the generator 
of this group are defined by relationship (8). Consequently, the 
manifold is fully identified by the differential equation 
$$i\frac {d\varphi}{dx}=m\varphi,\qquad m=\frac {2\pi}{l}.\eqno (9)$$
which is equivalent to this relationship.

Let us use this approach for the "topological" decoding of 
differential equation that are more complex than (9), namely, 
the equation for free Dirac field
$$i\gamma_1\frac{\partial}{\partial x_1}\psi-\sum_{\alpha =2}^4 
i\gamma_\alpha \frac
{\partial}{\partial x_\alpha} \psi=m\psi,\eqno (10)$$
which is obtained from Eq.(3) at $A_1=A_2=A_3=A_4=0$.

     To this end, let us again consider, for clearness, one of 
the simplest and well-studied two-dimensional closed connected 
nonorientable manifolds, namely, the Klein bottle, and encode its 
topological and metric properties using the differential 
equations within the framework of the same approach as was 
demonstrated by the example of a one-dimensional ring. It will be 
seen below that the solution of this problem and its 
generalization to the four-dimensional case allow the topological 
interpretation of the Dirac equation (10).
     The Klein bottle is obtained by gluing together two M\"obius 
strips along their edges. Euclidean plane is the covering surface 
for the Klein bottle, and the sliding symmetry group generated by 
two parallel translations and two 
reflections in the directions perpendicular to the translations 
is its fundamental group [8, 9]. Let us assume that two 
translation generators $l_1$ and $l_2$ satisfy the additional constraint
$$\frac {1}{l_1^2}+\frac {1}{l_2^2}=\frac {1}{l_0^2}, \eqno (11)$$
where $l_0$ is an additional metric characteristic of the manifold. 
Let us now express the topological characteristic (fundamental 
group) of the Klein bottle and the additional metric condition 
(11) in terms of an equation which plays the same role as does 
Eq. (9) for the ring.
     Similar to (7), the function
$$\varphi (x,y)=\exp \left(-2\pi i\frac {x}{l_1}-2\pi i \frac {y}{l_2} \right). 
\eqno (12)$$
is chosen as a basis function for the subgroup of two parallel 
translations along the $OX$ and $OY$ axes. A two-component spinor
$$\chi = \left(\begin{array}{c} \chi_1 \\ \chi_2 \end{array}\right).
\eqno (13)$$
is chosen as a basis for the subgroup of reflections 
perpendicular to the $OX$ and $OY$ axes. The reason for which the 
spinor basis is chosen for the reflections is that, 
as will be seen below, only with such a choice the resulting 
equation leads to metric condition (11).
      As above, the operators of type (6) form the representation 
of a translation group with basis functions (12). As for the 
reflections about the planes passing through $OX$ and $OY$, the Pauli 
matrices $\sigma_x$ and $\sigma_y$ [10]
$$\sigma_x=\left(\begin{array}{cc}0&1\\1&0\end{array}\right),\qquad
\sigma_y=\left(\begin{array}{cc}0&-i\\i&0\end{array}\right).\eqno (14)$$
form the representation of these reflections for the spinor basis 
(13).
     Thus, the functions
$$\psi (x,y)=\chi \varphi (x,y), \eqno (15)$$
are the basis functions for the fundamental group of Klein bottle 
(sliding symmetry), while the operators $P_x$ and $P_y$
$$P_x=\sigma_x T_{l_1x},\qquad P_y=\sigma_y T_{l_2y}.\eqno (16)$$
form the representation of sliding symmetry along the $OX$ and $OY$ 
axes.
     It is now straightforward to carry out direct calculation to 
check on that the equation expressing, similar to Eq. (9), the 
topological invariant (fundamental group) of the Klein bottle and 
satisfying metric constraint (10) on the elements of this group 
has the form
$$(P_x+P_y)\psi=\frac {1}{l_0}\psi.\eqno (17)$$
Indeed, one can easily verify that the insertion of Eqs. (12)--
(16) into Eq. (17) yields relationship (11). With allowance made 
for Eq. (16), Eq. (17) can be written in the form of a 
differential equation
$$i\sigma_x \frac {\partial \psi}{\partial x}+i\sigma_y \frac {\partial \psi}
{\partial y}=\mu \psi,\qquad \mu=\frac{1}{l_0}.\eqno (18)$$

     Let us compare this equation with Eq. (10) for a free Dirac 
field taking into account that the Dirac matrices $\gamma_i$ entering Eq. 
(10) form, in the four-dimensional pseudo-Euclidean space, the 
representation of reflections about three axes perpendicular to 
the $OX$ axis, provided that the Dirac bispinors form the basis of 
this representation [10]. This comparison demonstrates that Dirac 
equation (10) is the generalization of Eq. (18) for two-
dimensional Klein bottle to the four-dimensional pseudo-Euclidean 
Minkowski space. In this case, the role of metric constraint (10) 
is played by the energy conservation law
$$E^2-p_x^2-p_y^2-p_z^2=m^2,\eqno (19)$$
which provides, within the topological interpretation of the free 
Dirac field as a space--time manifold, the metric relationships 
between the translation subgroup generators,
$$\frac {1}{l_1^2}-\frac {1}{l_2^2}-\frac {1}{l_3^2}-\frac {1}{l_4^2}=
\frac {1}{l_0^2},\eqno (20)$$
where $l_0^{-1}=m,\quad l_1^{-1}=E,\quad l_2^{-1}=p_x,\quad l_3^{-1}=p_y,\quad 
l_4^{-1}=p_z.$

     So, Dirac equation (10) can be interpreted as a relation 
encoding the topological and metric properties of a closed 
nonorientable space--time manifold, whose analogue in the two-
dimensional case is provided by the Klein bottle. In this case, 
the Dirac spinors function as the basis functions of the 
fundamental group of a manifold, whose covering space is a 
pseudo-Euclidean Minkowski space. The energy, momentum, and mass 
of the Dirac field and the energy conservation law are related to 
the translation subgroup generators by relationships (20). 
Evidently, the foregoing consideration can be regarded as a 
"purely" topological derivation of the Dirac equation, i.e., as a 
derivation based on an assumption that the quantum object 
represents a certain space--time manifold, without invoking the 
Lagrangian, Hamiltonian, or any other mechanical formalism.
     In closing this section, note that the closeness of a 
manifold in the pseudo-Euclidean space does not imply any 
constraints on its extension over the time axis. For example, a 
circle in the pseudo-Euclidean plane is mapped into an 
equilateral hyperbola in the usual plane [11].
\par\medskip

\noindent {\bf Geometrisation of electromagnetic interaction}
\par\medskip

     Let us now turn to the fundamental problem of justifying the 
interpretation of Eqs. (3)--(5) for the interacting 
electromagnetic field and its sources as group--theoretic 
relations that encode the topological characteristics and metric 
constraints of a certain unified nonmetrized space--time 4-
manifold.
     By now, the topology of 4-manifolds is understood to a much 
lesser extent than the topology of, e.g., two-dimensional 
manifolds (see, e.g., [12, 13]). The latter are classified in 
detail, and the parameters of their main topological groups are 
known [6--9]. For this reason, I will attempt to invoke a 
possible analogy with similar problem in the two-dimensional 
topology (much as the Dirac equation was derived topologically in 
the preceding section), because the successful use of "low-
dimensional" analogies is one of the merits of the geometrization 
of physical and mathematical problems. 

Specifically, let us find 
out what will happen if the different topological properties of 
the orientable and nonorientable manifolds are combined within a 
single unified two-dimensional closed manifold. Consider, for 
example, what is a hybrid of torus and Klein bottle like and how 
can its topological properties be described.
     According to the topological classification of two-
dimensional manifolds, torus is a "sphere with one arm" 
(orientable surface of the genus $p = 1$) and the Klein bottle is a 
"sphere with two holes glued up by M\"obius strip" (nonorientable 
surface of the genus $q = 2$) [6--9]. By the hybrid of torus and 
Klein bottle can be meant a sphere to which one arm and two 
M\"obius strips are glued simultaneously. Such a surface is 
nonorientable and belongs to the genus $q' = 2p+q = 2 + 1 = 3$ type [6--9]. 
For $q > 2$, a 
hyperbolic plane is a universal covering surface of a manifold 
and the fundamental group is generated by $q$ sliding symmetries 
[8, 14].
     If one assumes that the analogy with the manifold of the 
above-mentioned type "operates" in our case, one should expect 
that Dirac equation (1) can be interpreted as a metric relation 
for the nonorientable 4-manifold, whose fundamental group is 
generated by sliding symmetries, while the universal covering 
space is a four-dimensional pseudo-Euclidean analogue of the 
hyperbolic plane, namely, a space with semimetric transfer [19] or conformally 
pseudo-Euclidean space [11].

Let us show that the Dirac equation indeed allows the above 
interpretation and consider for simplicity conformally Euclidean space. 
Conformally Euclidean space is said to be a 
Riemannian space that can be conformally mapped into the 
Euclidean space. By this is meant that every point $M(x)$ of the 
conformally Euclidean space can be assigned a point $M_E$ in the 
Euclidean space so that the corresponding differentials of arc 
lengths are related to each other at every point by the 
relationship [11]
$$ds^2_E=f(x^0,x^1,x^2,x^3)ds^2, \eqno (21)$$
where $f(x)$ is a certain function of coordinates, $ds^2=g_{ik}dx^idx^k$ 
defines the metric of the conformally Euclidean space, and 
$ds^2_E=g^E_{ik}dx^idx^k$ is the squared 
differential arc length in the Euclidean space (in our case of 
pseudo-Euclidean space, $g_{00}^E=1, g_{11}^E=g_{22}^E=g_{33}^E=-1$ and
$g_{ik}^E=0, i\ne k$).

Let us now turn to the left-hand side of Dirac equation (3). 
Compared to equation (10) for a free electron--positron field, it 
includes the expression $(\partial /\partial x_l-ieA_l)$ instead of the 
usual derivative $\partial /\partial x_l$. In 
electrodynamics, this expression is customarily called "covariant 
derivative", because it formally resembles the covariant 
derivative $\bigtriangledown_l$ of a covariant vector field $B_m$ 
and is written as [6, 11]
$$\nabla_l B_m=\frac{\partial B_m}{\partial x_l}+\Gamma^s_{ml}B_s,
\eqno (22)$$
where $\Gamma^s_{ml}$ is the connectivity.
     The geometrical meaning of a connectivity is, in particular, 
that the covariant derivative plays the role of a translation 
group generator for the usual (not spinor) tensor fields on a 
manifold [6, 11]. (In the Euclidean space, the connectivity is 
zero and the "usual" derivative $\partial /\partial x_l$ 
plays the role of a translation 
group generator). However, for the spin-tensor fields and, in 
particular, for the four-component first-rank spin tensor 
entering the Dirac equation, the connectivity with the above
mentioned properties does not exist in an arbitrary Riemannian 
space. This is caused by the fact that spin tensors are Euclidean 
rather than affine, and the transformation law for their 
components is specified by the rotation group representation, 
which cannot be continued to a group of all linear 
transformations [10, 11]. In other words, the spin tensors can be 
connected to each other at different points of a curved space 
only if the orthonormalized frame remains orthonormalized 
upon the translation in this space.

However, for a particular case of Riemannian spaces, namely, 
for the conformally Euclidean spaces, a certain vicinity of an 
arbitrary point $M$ can always be mapped, with retention of metric, 
into the vicinity of the other arbitrary point $M'$, and this can 
be done in such a way that the orthonormalized frame 
specified at point $M$ transforms into an arbitrarily chosen 
orthonormalized frame at $M'$ [11]. For this reason, the 
parallel translation of spinor in such a space is determined by 
metric from the same formulas as occur for usual tensors, with 
only the $\Gamma^p_{lp}$-type connectivity components being nonzero. 
(Weil was, probably, the first to realize it [15, 16]).
     Recall that not the physical space of events is assumed to 
be a conformally Euclidean space in the approach suggested in 
this work but a universal covering space, which serves only as a 
mathematical tool for describing the fundamental group of a 
physical object, namely, of a closed connected 4-manifold of 
interacting electromagnetic and electron--positron fields. 

Let us 
now assume that the expression $ieA_l$ in Eq. (3) can be regarded as a 
connectivity $\Gamma^p_{lp}$ in the conformally Euclidean space. Then the 
expression $(\partial /\partial x_l - ieA_l)$ 
entering in Dirac equation (3) can be interpreted 
as a translation group generator along the universal covering 
surface of a certain connected 4-manifold. The translation group 
generator multiplied by the symmetry operator $\gamma_l$ about the 
hyperplanes passing through the $OX_l$ axis can be regarded as a 
generator of a "local" sliding symmetry group in the spinor basis. 
This leads to the main conclusion of this work: Dirac equation 
(3) can be interpreted as a group-theoretic relation describing 
the metric and topological properties of a certain 4-manifold, 
for which a space  with semimetric transfer is a universal 
covering surface, while a group determined by the local sliding 
symmetry is the fundamental group.

The fact that $ieA_l$ in the above interpretation has a meaning of 
connectivity on the covering space also allows the geometrical 
interpretation of the electric and magnetic field components, 
i.e., of the components of the electric and magnetic 
fields tensor $F_{ik}$. To this end, let us use the fact that the curvature 
tensor $R^q_{lk,i}$ (Riemann--Christoffel tensor defining the deviation from 
Euclidean geometry in the Riemannian space) is expressed in terms 
of connectivity as [6, 11]
$$R^q_{lk,i}=\left(\frac{\partial \Gamma^q_{li}}{\partial x_k}-
\frac{\partial \Gamma^q_{ki}}{\partial x_l}+\Gamma^q_{kp} \Gamma^p_{li}
-\Gamma^q_{lp} \Gamma^p_{ki}\right).\eqno (23)$$
(As before, the summation over repeating indices goes from $0$ to $3$).
     Let us contract the curvature tensor with respect to its 
upper and lower right indices (the resulting tensor is denoted by 
$R^0_{lk}$),
$$R^0_{lk}=R^q_{lk,q}=\frac{\partial \Gamma^q_{lq}}{\partial x_k}-
\frac{\partial \Gamma^q_{kq}}{\partial x_l}.\eqno (24)$$

     Comparing Eq. (24) with Eq. (4) and using the fact that $\Gamma^q_{mq} = 
ieA_m$, one obtains
$$ieF_{ik}=R^0_{ik},\eqno (25)$$
i.e., within the geometrical interpretation, the tensor of 
electric and magnetic fields coincides, except for the factor 
$ie$, with certain components of the curvature tensor of a 
universal covering surface. Therefore, Maxwell Eq. (5) relates 
the above-mentioned components of curvature tensor to the basis 
functions of the fundamental group, thereby rendering the system 
of Eqs. (3)--(5) closed. The curvature tensor for a space with 
constant curvature $K$ has the form [11]
$$R_{ij,kl}=K(g_{ik}g_{jl}-g_{il}g{jk}).\eqno (26)$$
Comparing Eqs. (26) and (25), one arrives at the conclusion that, 
within the geometrical interpretation, the electric charge $e$ is 
proportional to a constant curvature $K$.

Thus, in the geometrical interpretation, the equations of 
classical relativistic electrodynamics (3)--(5) have the form
$$i\gamma_l(\frac{\partial}{\partial x_l}-\Gamma^p_{lp})\psi=m\psi,\eqno (27)$$
$$R^0_{ik}=\frac{\partial \Gamma^p_{ip}}{\partial x_k}-\frac{\partial
\Gamma^p_{kp}}{\partial x_i},\eqno (28)$$
$$\sum_{i=1}^4 \frac{\partial R^0_{ik}}{\partial x_i}=
ie^2\psi*\gamma_1\gamma_k\psi.\eqno (29)$$
\par\medskip

\noindent {\bf Weak interaction}
\par\medskip

     Let us now show that, in a one-particle approximation 
adopted in this work (low energies), weak interaction can be 
represented as a manifestation of the torsion in the covering 
space of a 4-manifold representing weak field and its sources.
     In due time, Einstein attempted at including electromagnetic 
field into a unified geometrical description of physical fields 
by "adding" torsion to the Riemannian space--time curvature, 
which reflects the presence of a gravitational field in general 
relativity [17]. Since the curvature of covering space 
corresponds within our approach to the electromagnetic field, I will attempt to 
include weak interaction into the topological approach by 
including torsion in this space.

It is known that weak interaction breaks the mirror space-time symmetry. On
the other hand above symmetry can be violated in the space with torsion
(left screw looks like a right one in the mirror). So
it is natural to assume that within our topological approach 
torsion may be connected with weak interaction. Let us 
first consider the case where the electromagnetic field is 
absent, i.e., the curvature of covering space is zero.
     A space with torsion but without curvature is called the 
space with absolute parallelism [11]. Thus, the challenge is to 
determine how does free-particle Eq. (10) change if the 
interparticle interaction is due only to the torsion, which 
transforms the pseudo-Euclidean covering space into a space with 
absolute parallelism. 

Let us denote the torsion tensor by $S^k_{lm}$; then 
the problem can be formulated as follows. It is necessary to 
"insert" the tensor $S^k_{lm}$ or some of its components into Eq. (10) so 
that the resulting equation remains invariant about the Lorentz 
transformations and adequately describes the experimental data 
(e.g., violation of spatial and time symmetry by weak 
interaction).
     Among the spaces with torsion, there are so-called spaces 
with semisymmetric parallel translation [18,19]. The torsion tensor 
$S^k_{lm}$ for such spaces is defined by the antisymmetric part of 
connectivity and can be represented in the form
$$S^k_{lm}=S_lA^k_m-S_mA^k_l.\eqno (30)$$
Here, $S_l$ is a certain vector and $A^k_l$ is the identity tensor. 
The vector $S_l$ has the property that the 
infinitesimal parallelogram remains closed upon parallel 
translation in the hyperplanes perpendicular to this vector. 

One 
may thus assume that in the presence of vector $S_l$ the spatial 
isotropy breaks in such a way that the isotropy is retained only 
in the indicated hyperplanes.
     Assuming that along $S_l$ only the translational symmetry is retained, 
while the symmetry of Eq. (10) is retained in the hyperplane 
perpendicular to this vector, one can recast this equation as
$$i\left (\frac{\partial}{\partial X_1}-S \right)\varphi-
\sum^4_{k=2}i\sigma_k\frac
{\partial \varphi}{\partial X_k}=m\varphi.\eqno (31)$$
Here, the $X_1$ axis is aligned with the vector $S_l$, $\varphi$ is a 
two-component spiner, $\sigma_k$ denotes a 
two-row matrices of rotation group representation in the spinor 
basis (Pauli matrices) and $S$ is some tensor connection in considering
space.
     
The question of how does 
Eq. (31) account for the other properties of weak interaction and 
how does it relate to the results of the standard model of 
electroweak interaction will be considered in detail elsewhere. Notice
only that $U(1)SU(2)$--gauge transformation looks here as
production of the S-vector rotation and rotation
within the 3-dimentional hyperplane perpendicular to this vector.

\par\medskip
\noindent
{\bf References}
\par\medskip
\noindent 1. O.A.Olkhov, Proceedings of the 7th International 
Symposium on Particles, Strings and
Cosmology, Lake Tahoe, California, 10-16 December 1999. Singapure-New Jersey-
Hong-Kong, World Scientific, 2000, p.160, quant-ph/0101137;
Chem.Phys. 2000. V.19. N 5. p.92.; V.19. N 6. p.13,(in russian);
Thesises of the International Seminar on Physics of Electronic and Atomic
Collisions, Klyazma, Moscow region, Russia, 12-16 March 2001, p.28,
e-print quant-ph/0103089\\
2. V.Heine, Group theory in quantum mechanics. Pergamon Press. London-
Oxford-N.Y.-Paris, 1960\\
3. V.B.Berestetzki, E.M.Lifshitz, L.P.Pitaevski, Relyativistskay kvantovaya 
teoriya, Moscow, Nauka, 1968\\
4. I.S.Bell, Rev.Mod.Phys. 38(1966)447\\
5. von J.V.Neumann., Mathematische grundlagen der quantenmechanik,
Verlag von Julius\\ Springer, Berlin, 1932\\
6. B.A.Dubrovin, S.P.Novikov, A.T.Fomenko, Sovremennay geometriay, V.2, Ch.4,
Moscow, Editorial, 2001\\
7. Yu.G.Borisovitch, N.M.Bliznaykov, Ya.A.Izrailevitch, T.N.Fomenko,
Vvedenie v topologiyu, Ch.3, Moscow, Nauka, 1995\\
8. H.S.M.Coxeter, Introduction to geometry. John Wiley and Sons, 
N-Y-London, 1961\\
9. A.T.Fomenko, Naglyadnay geometriay i topologiay, Ch.2, Moscow, 
Moscow University, 1998\\                    
10. E.Cartan, Le\c{c}ons sur la th\'eorie des spineurs. Paris, 1938\\
11. P.K.Raschevski, Rimanova geometriay i tensornii analiz, Ch.3, Moscow,
Izdatelstvo techniko-teotetitcheskoi literaturi, 1953\\
12. M.Freedman, J.Diff.Geometry. 17(1982)357\\
13. A.T.Fomenko, Diferentcialnay geometriay i tipologiay. Dopolnitelnie
glavi, Izevsk, Izevskay respublikanskay tipografiay, 1999\\
14. H.S.M.Coxeter, W.O.Moser, Generators and Relations for discrete
groups. New York-London.: John Wiley and Sons, 1980\\
15. H.Weyl, Gravitation und Electrizit\"at. Berlin, Sitzber.:
Preuss. Akad. Wiss, 1918\\
16. H.Weyl, Z.f.Phys. 56(1929)330\\
17. A.Einstein, Riemann Geometrie mit Aufrechterhaltung des 
Fernparallelismus.
Sitzungsber: preuss. Akad. Wiss., phys-math. K1., 1928, 217-221\\
18. J.A.Schouten, Tensor analysis for physicists. Oxford, 1952\\
19. J.A.Schouten and D.J.Struik, Einf\"uhrung in die neueren Methoden der
Differentialgeometrie, I. Groningen--Batavia.: Noordhoff, 1935

\end{document}